\documentclass{appolb}
\usepackage{graphicx}

\def\Pom{{\bf I\!P}}

\begin{document}
\title{Double scattering production of two $\rho^0$ mesons in UPC%
\thanks{Presented at the 16th conference on Elastic and Diffractive scattering, EDS Blois 2015 }%
}
\author{Mariola K{\l}usek-Gawenda, Antoni Szczurek
\address{H.Niewodnicza\'nski Institute of Nuclear Physics, Polish Academy of Sciences, Radzikowskiego 152, 31-342 Krak\'ow, Poland \and University of Rzesz\'ow, Rejtana 16, 35-959 Rzesz\'ow, Poland}
}
\maketitle
\begin{abstract}
We review our recent results for double-scattering
mechanism in the exclusive $AA \to AA \rho^0\rho^0$ reaction
in ultrarelativistic ultraperipheral heavy ion collisions. 
The cross section for single and double-$\rho^0$ production
is calculated in the impact parameter space 
equivalent photon approximation.
We compare the results of our calculation
with the STAR and ALICE Collaboration results
on one $\rho^0(770)$ meson and four-charged-pion production.
\end{abstract}

\section{Introduction}

The exclusive production of simple final states
in ultrarelativistic UPC of heavy ions
is a special class of nuclear reactions \cite{reviews}.
At high energies and due to large charges of colliding nuclei
(gold-gold or lead-lead collisions) there are 
two categories of the underlying reaction mechanisms.
First is the photon-photon fusion \cite{gamgam_rho0rho0} 
and the second is double photoproduction \cite{DS_2014}.

We evaluated for the first time differential distributions 
for exclusive production of two $\rho^0(770)$ mesons 
(four charged pions) in the double scattering (photon-Pomeron) process
\cite{DS_2014} for RHIC and LHC.
The results will be compared with the contribution 
of two-photon mechanism. 
The analysis includes a smearing of $\rho^0$ mass 
using a parametrization of the ALICE Collaboration \cite{rho_smearing}.

\section{Single-scattering mechanism}

\begin{figure}[htb]
\centerline{%
\includegraphics[scale=0.3]{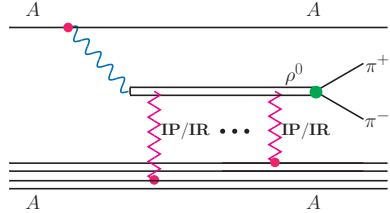}}
\caption{Single vector meson production by photon-Pomeron (or Pomeron-photon)
fusion.}
\label{Fig:gIP_rho0}
\end{figure}

Fig. \ref{Fig:gIP_rho0} illustrates a single $\rho^0$ production mechanism
(and its decay into $\pi^+\pi^-$ state) in UPC of heavy ions.
Photon emitted from a nucleus fluctuates into hadronic or quark-antiquark
components and converts into an on-shell meson.
The cross section for this mechanism can be written differentially
in the impact parameter $b$ (distance between two colliding nuclei)
and in the vector meson rapidity $y$
\begin{equation}
\frac{\mathrm{d} \sigma_{AA \to AA \rho^0}}{\mathrm{d}^2b \mathrm{d}y}
= \frac{\mathrm{d} P_{\gamma \Pom} (b,y)}{\mathrm{d}y} 
+ \frac{\mathrm{d} P_{\Pom \gamma} (b,y)}{\mathrm{d}y} \;.
\label{eq:sigma_single}
\end{equation}
$P_{\gamma \Pom / \gamma \Pom} (b,y)$ is the probability density
for producing a vector meson at rapidity $y$ for fixed impact parameter $b$
of the heavy ion collision. Each probability is the convolution
of the $\gamma A \to \rho^0 A$ cross section
 and a flux of equivalent photon
\begin{equation}
P_{\gamma \Pom / \gamma \Pom} (b,y) =
\omega_{1/2} \, N(\omega_{1/2},b) \, \sigma_{\gamma A_{2/1} \to \rho^0 A_{2/1}} \;,  
\end{equation}
where 
$N(\omega_{1/2},b)$ is usually written as a function of 
the impact parameter $b$.
The photon flux is expressed through nuclear form factor $F(q)$
which is related to charge distribution in the nucleus.
Details of different form factors and their application
to nuclear calculation can be found 
in Refs. \cite{gamgam_rho0rho0,muons,Jpsi}.
To calculate the $\sigma_{\gamma A_{2/1} \to \rho^0 A_{2/1}}$
cross section we use a sequence of equations which are presented
in \cite{KN1999}. Constants for the underlying
$\sigma_{\gamma p \to \rho^0 p}$ cross section are obtained
from a fit to HERA data \cite{HERA}.
The $\sigma_{\rho^0 A}$ total cross section 
can be calculated using either classical mechanics
\begin{equation}
\sigma_{\rho^0 A} = \int \mathrm{d}^2 \textbf{r} 
\left( 1-\exp \left( -\sigma_{\rho^0 p} \, T_A\left( \textbf{r} \right) \right) \right)
\label{eq:CM}
\end{equation}
or quantum mechanical Glauber formula
\begin{equation}
\sigma_{\rho^0 A} = 2 \int \mathrm{d}^2 \textbf{r} 
\left( 1-\exp \left( - \frac{1}{2} \sigma_{\rho^0 p} \, T_A\left(
      \textbf{r} \right) \right) \right) \; ,
\label{eq:QM}
\end{equation}
where $r$ is a distance between photon emitted from first/second nucleus 
and middle of second/first one. 
$T_A\left( \textbf{r} \right)$ is the nuclear thickness function.

Returning to formula for single vector meson production
(Eq. \ref{eq:sigma_single}), this expression depends on the running
$\rho^0$ meson mass. In our calculation we use the ALICE parametrization
\cite{rho_smearing} which is the most appropriate for the LHC data
(a comparison of the ZEUS, STAR and ALICE parameters
for relativistic Breit-Wigner and continuum amplitudes was shown
during a talk at the EDS Blois workshop and can be found in \cite{MKG_thesis}).

\begin{figure}[htb]
\centerline{
\includegraphics[scale=0.26]{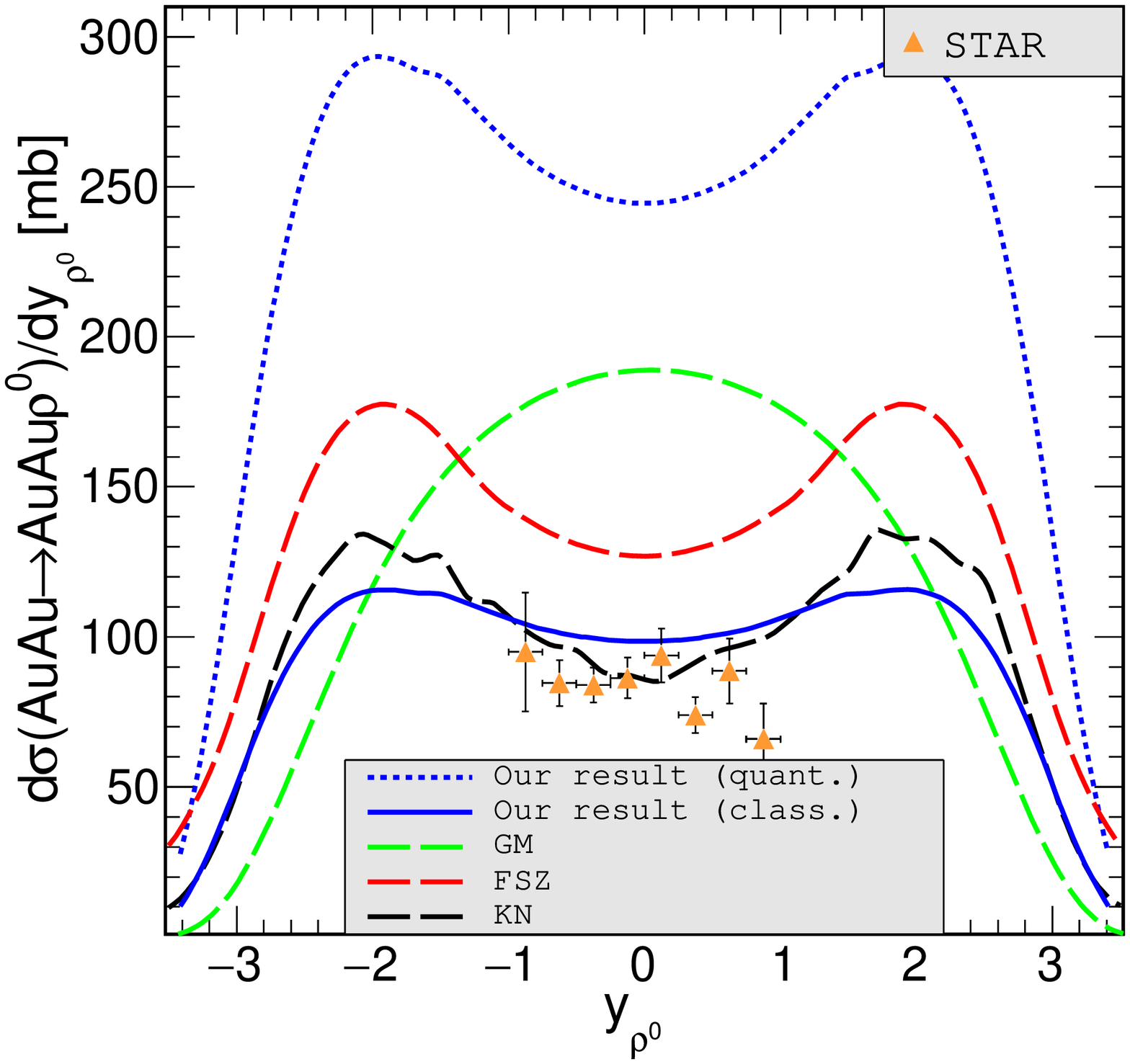}
\includegraphics[scale=0.26]{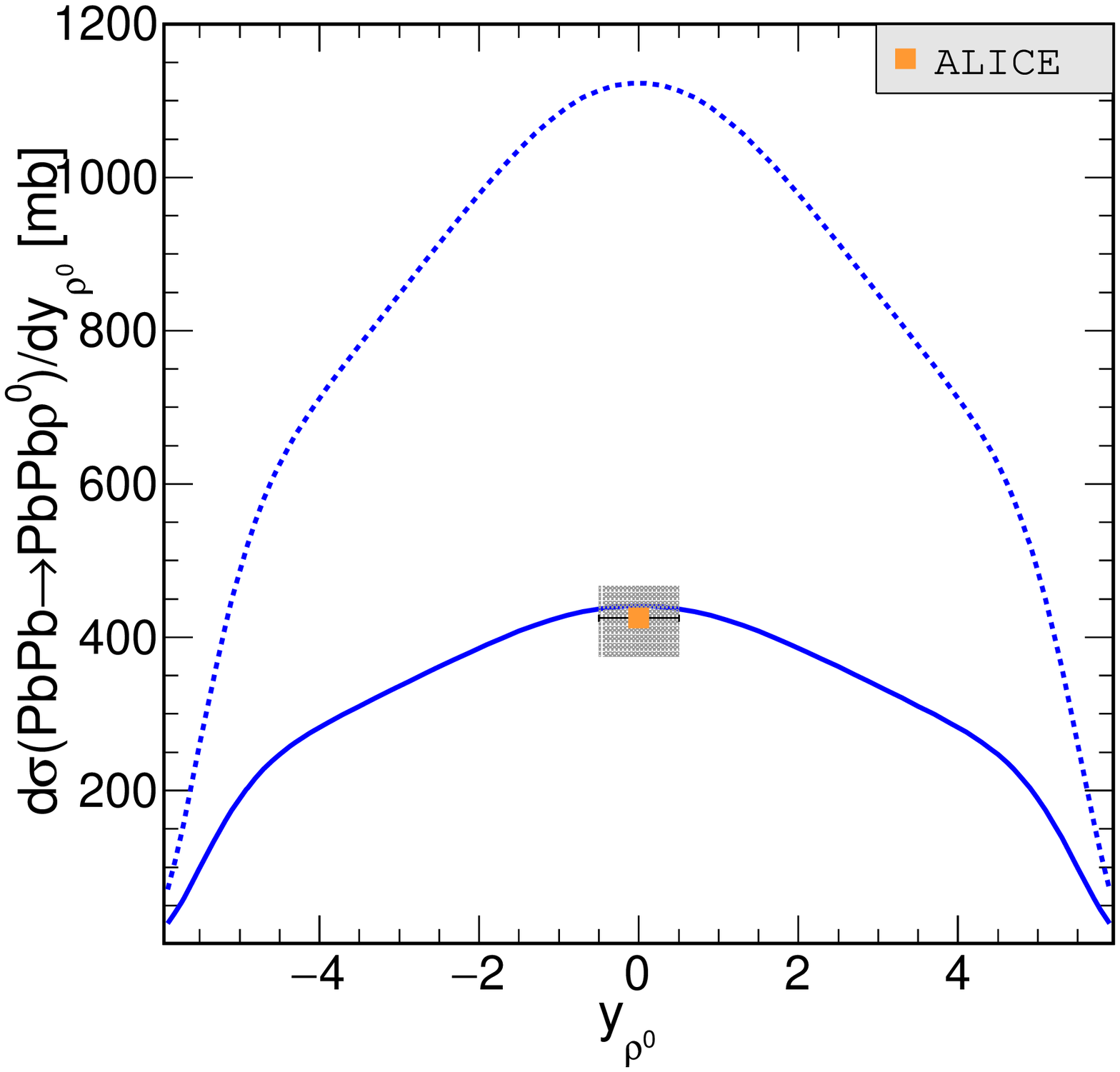}
}
\caption{Rapidity distribution of $\rho^0$ meson with the smeared mass 
of the meson
for gold-gold collisions at the RHIC energy (left panel) and
for lead-lead collisions at the LHC energy (right panel).}
\label{Fig:dsig_dy_single}
\end{figure}

Fig. \ref{Fig:dsig_dy_single} shows the comparison of cross section
for coherent $\rho^0$ production measured by the STAR \cite{STAR} (left panel)
and ALICE \cite{rho_smearing} (right panel) Collaborations
for different theoretical models (\cite{KN1999,GM, FSZ}).
In addition, one can observe that calculations for classical  
rescattering (Eq. \ref{eq:CM}) better (than in quantum approach 
(Eq. \ref{eq:QM})) 
describe both the STAR and ALICE experimental data.
Our results relatively well describe the STAR and ALICE experimental data
for the single vector meson photoproduction in heavy ion UPC.
This fact is important for calculation of cross section for double-scattering
mechanism which is discussed in the next section.

\section{Double $\rho^0$ production}

\begin{figure}[htb]
\centerline{%
\includegraphics[scale=0.35]{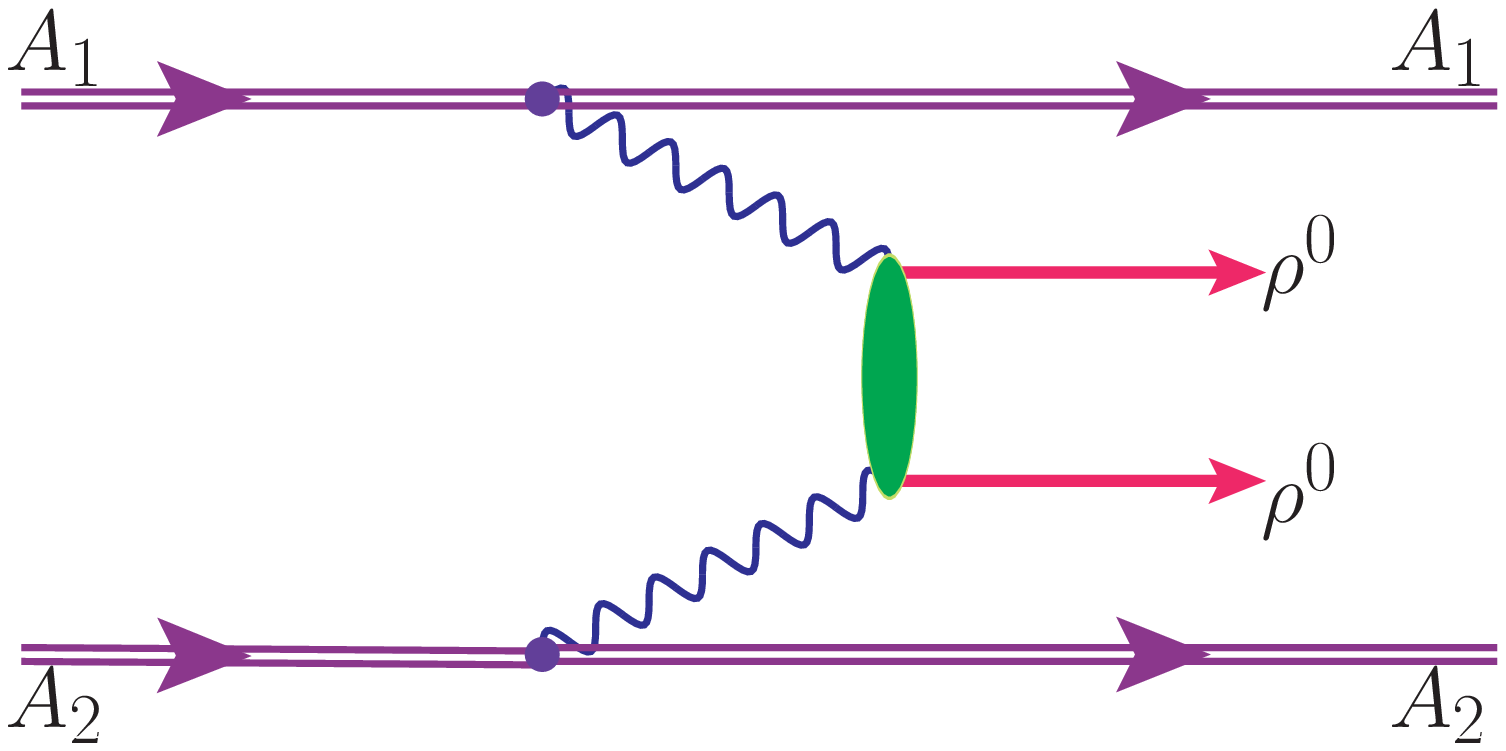}
\includegraphics[scale=0.3]{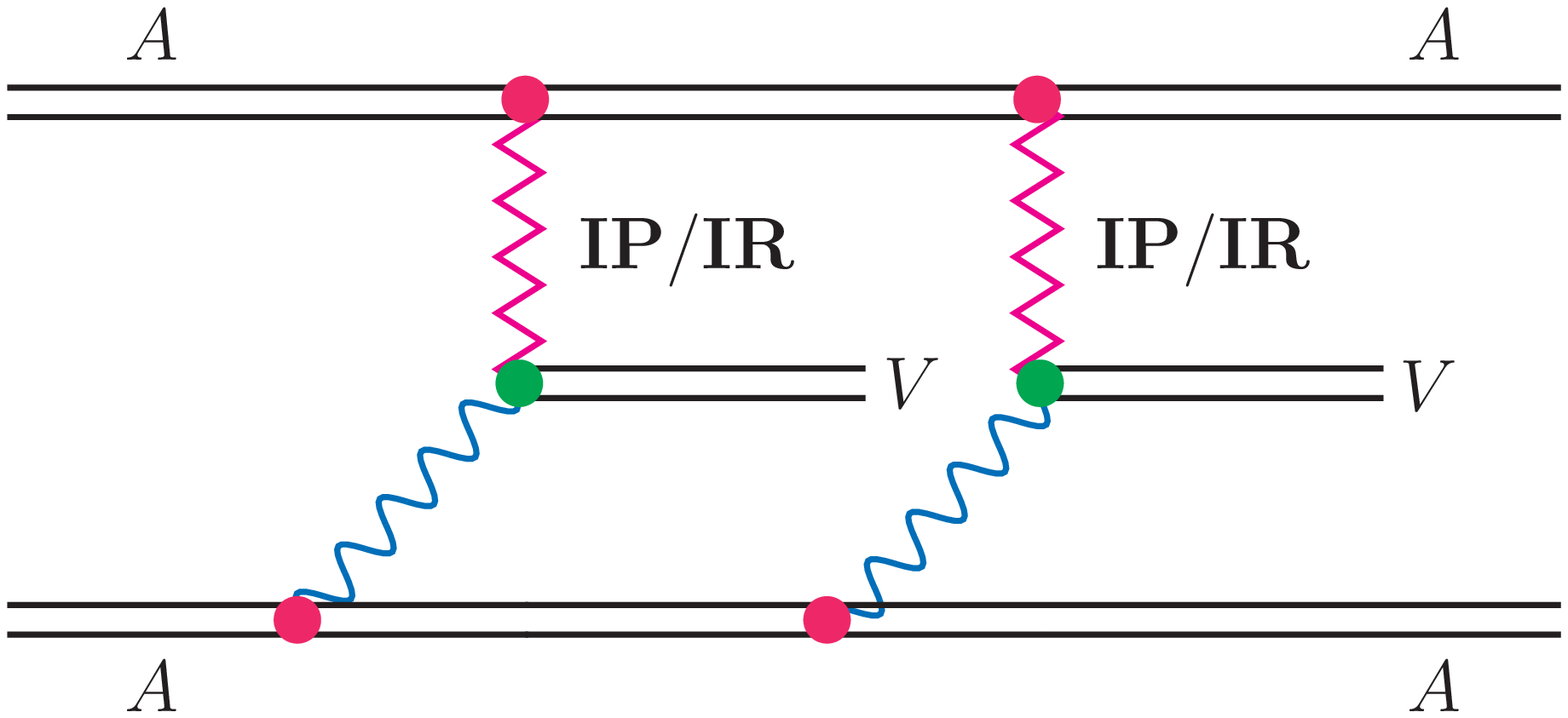}}
\caption{Double $\rho^0$ production for $\gamma\gamma$ fusion (left panel)
and for double-scattering mechanism (right panel).}
\label{Fig:DS}
\end{figure}

Fig. \ref{Fig:DS} shows diagrams for production of pairs of vector mesons
by two-photon-induced subprocess in heavy ion collision (left panel)
and for the double-scattering mechanism (in fact, we take into account
four different combinations of $\gamma \Pom$ exchanges: 
$\gamma \Pom-\gamma \Pom$, $\gamma \Pom - \Pom \gamma$,
$\Pom \gamma - \Pom \gamma$ and $\Pom \gamma - \gamma \Pom$).

The cross section for exclusive $\rho^0\rho^0$ production via
$\gamma\gamma$ fusion is closely explained 
in Ref. \cite{gamgam_rho0rho0}.
There the elementary cross section ($\gamma \gamma \to \rho^0\rho^0$) 
is divided into two parts: a low-energy component 
($W_{\gamma\gamma}=(1-2)$ GeV)
and a VDM-Regge parametrization ($W_{\gamma\gamma}>2$ GeV)
\cite{gamgam_rho0rho0}.
The cross section for the double $\rho^0$ photoproduction
is expressed with the help of probability density
of single $\rho^0$ meson production as
\begin{eqnarray}
\frac{\mathrm{d} \sigma_{AA \to AA \rho^0 \rho^0}}{ \mathrm{d}y_1 \mathrm{d}y_2}
= \frac{1}{2} & \int&  
\left( \frac{\mathrm{d} P_{\gamma \Pom} (b,y_1)}{\mathrm{d}y_1} 
+ \frac{\mathrm{d} P_{\Pom \gamma} (b,y_1)}{\mathrm{d}y_1} \right) \nonumber \\
& \times& 
\left( \frac{\mathrm{d} P_{\gamma \Pom} (b,y_2)}{\mathrm{d}y_2} 
+ \frac{\mathrm{d} P_{\Pom \gamma} (b,y_2)}{\mathrm{d}y_2} \right)
\mathrm{d}^2b \;.
\label{eq:sigma_double}
\end{eqnarray}
The factor $\frac{1}{2}$ appears due to identity of mesons in the outgoing channel.
\begin{figure}[htb]
\centerline{
\includegraphics[scale=0.26]{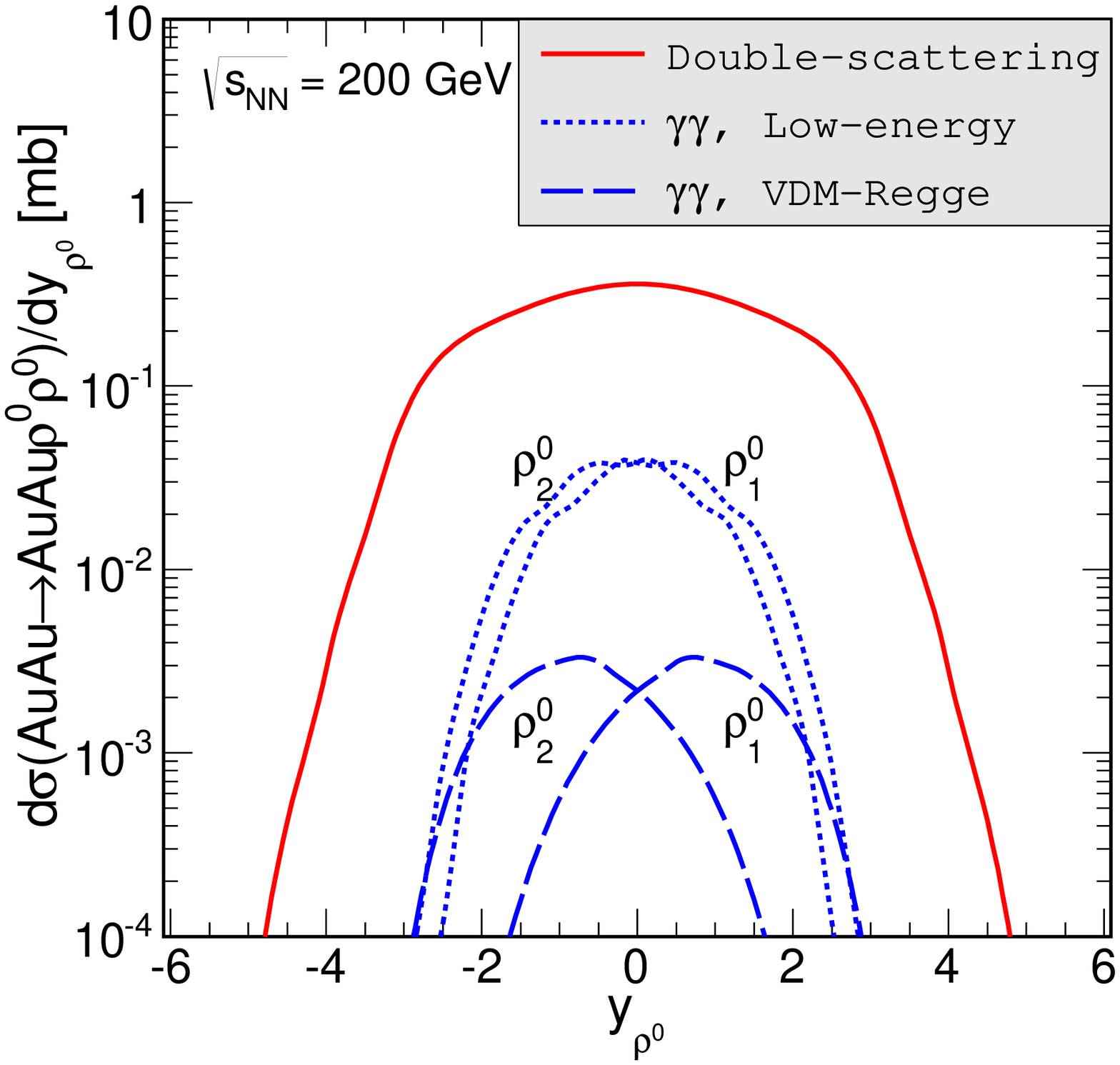}
\includegraphics[scale=0.26]{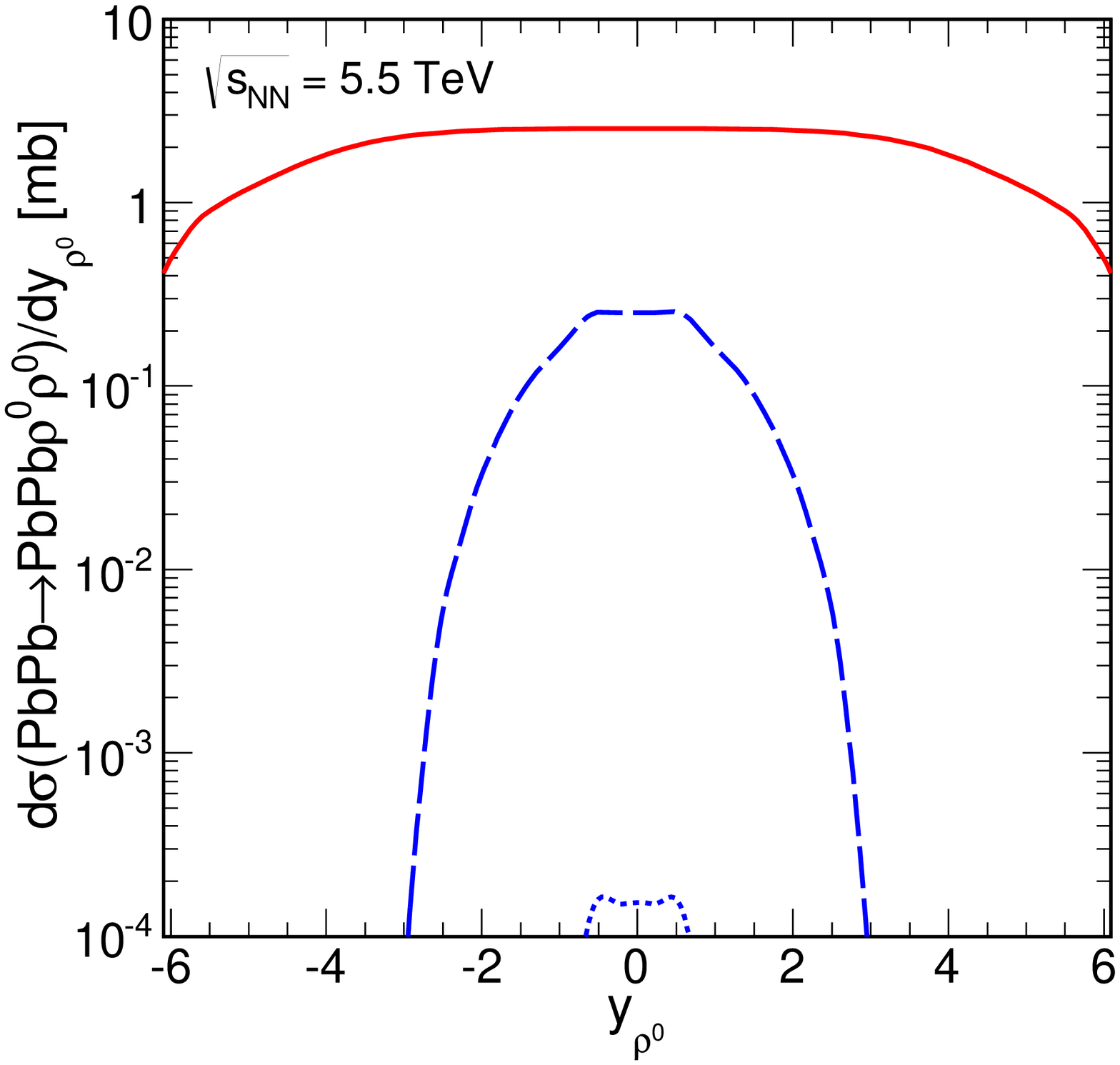}
}
\caption{Rapidity distribution of one of the $\rho^0$ meson produced in the double-scattering mechanism and in the $\gamma\gamma$ fusion at the RHIC (left panel)
and at the LHC (right panel) energy.}
\label{Fig:dsig_dy_double}
\end{figure}

Fig. \ref{Fig:dsig_dy_double} shows differential cross section as
a function of one $\rho^0$ meson rapidity and a comparison of 
the double-scattering and $\gamma\gamma$ fusion mechanisms at 
RHIC (left panel) and at LHC (right panel) energy. 
One can observe a clear dominance of the DS component 
over the $\gamma\gamma$ component. The distribution for the center 
of mass energy $\sqrt{s_{NN}}=5.5$ TeV is much broader than that for 
$\sqrt{s_{NN}}=200$ GeV.
At the LHC energy the higher values of two-meson invariant mass 
becomes more important which corresponds to larger values of particle
rapidity. Thus the high-energy component of the elementary cross section
dominates at the LHC energy. Somewhat surprisingly at this energy, 
the VDM-Regge component is about three orders of magnitude larger than 
the low-energy component, which is opposite to the case of the RHIC energy. 
Both at the RHIC and LHC energy, 
the contributions coming from the $\gamma\gamma$ fusion is one order 
of magnitude smaller than that from the double-scattering mechanism. 

\begin{figure}[htb]
\centerline{
\includegraphics[scale=0.26]{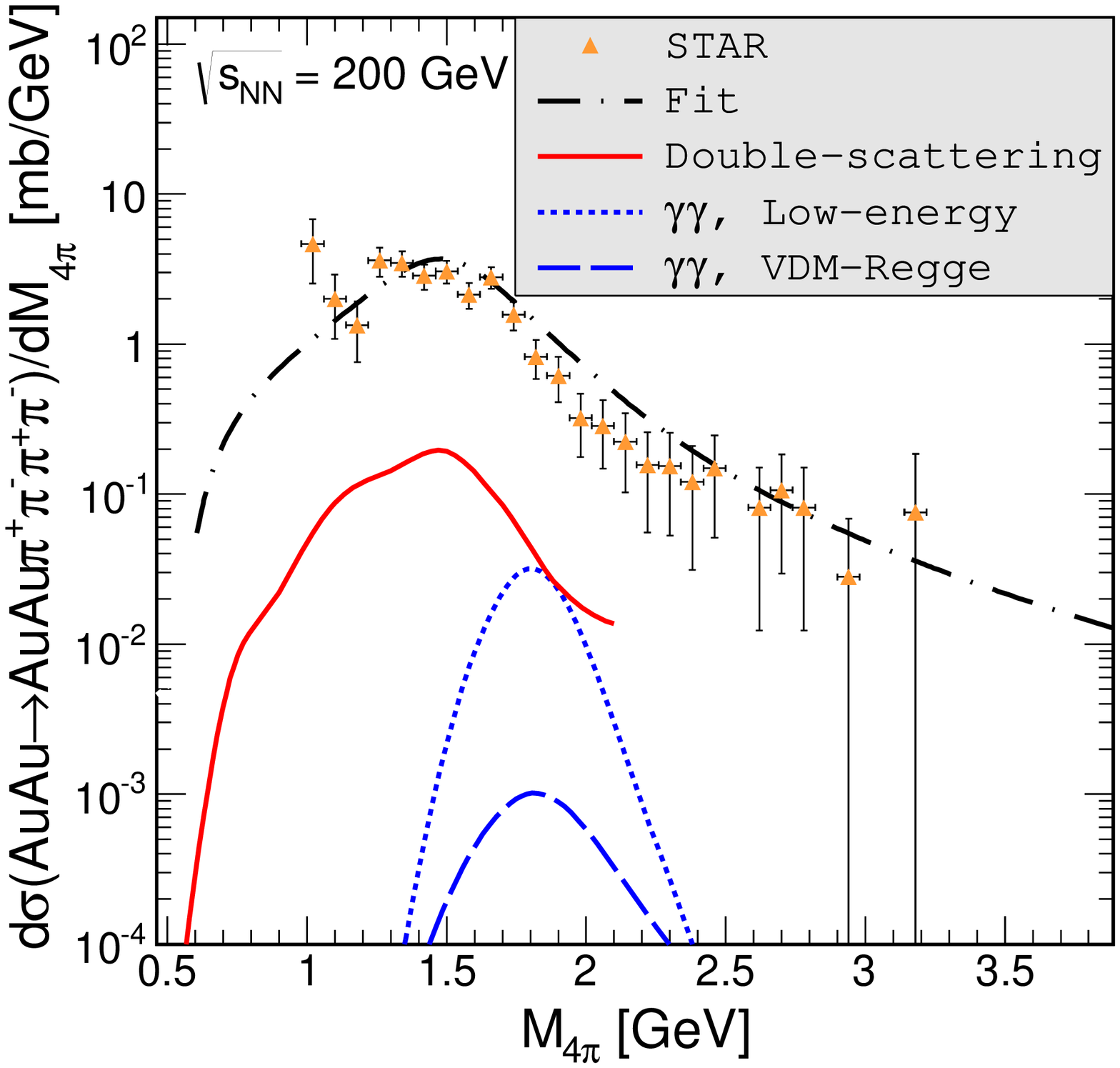}
\includegraphics[scale=0.26]{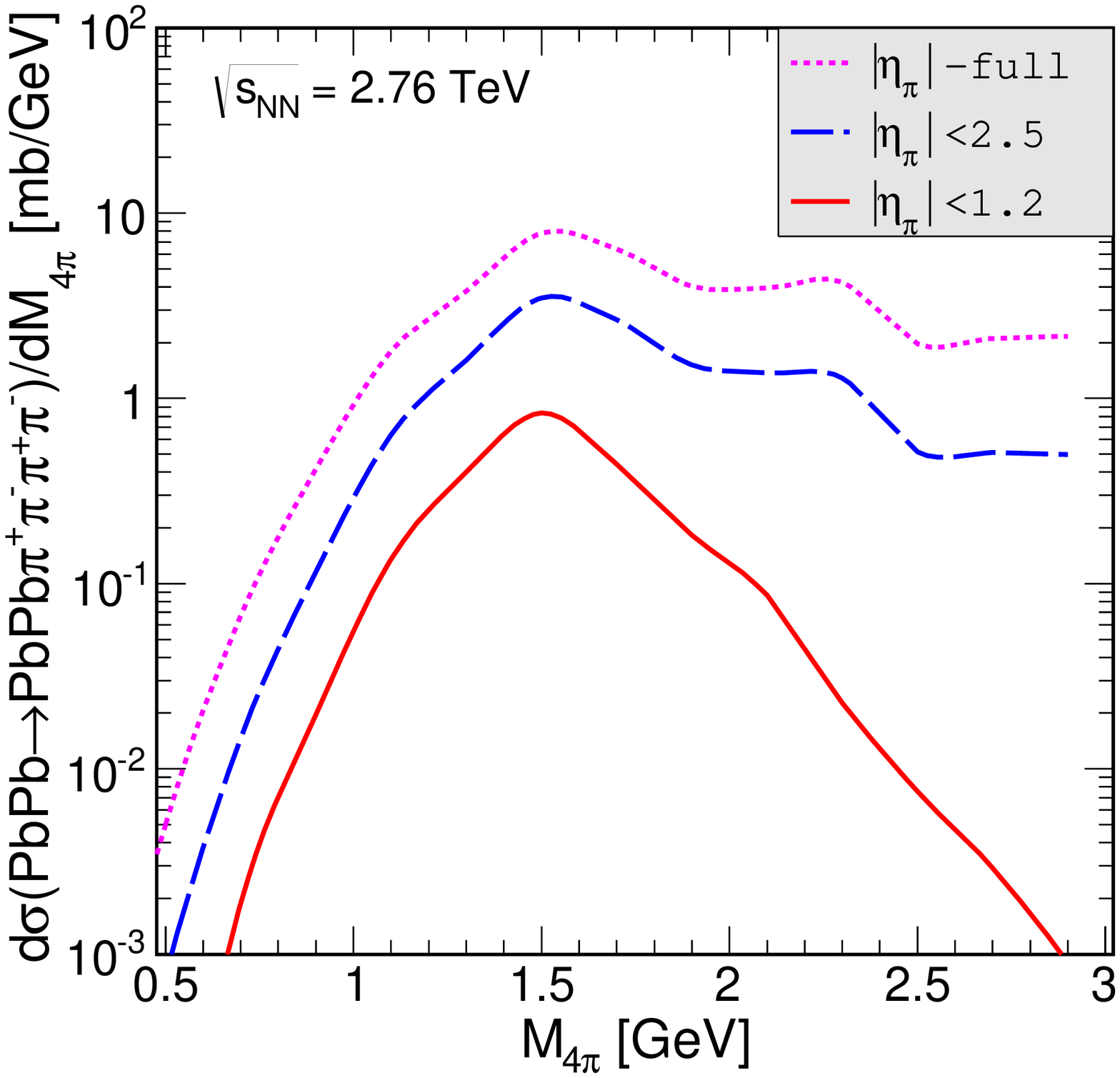}
}
\caption{Four-pion invariant mass distribution for the limited acceptance
of the STAR experiment (left panel)
and for the limited range of pion pseudorapidity at the LHC energy 
(right panel).}
\label{Fig:dsig_dM4pi}
\end{figure}

The left panel of Fig. \ref{Fig:dsig_dM4pi} shows four-pion invariant mass 
distribution for double-scattering, low-energy bump and high-energy VDM-Regge 
$\gamma\gamma$ fusion mechanism for the limited acceptance of the STAR 
experiment ($|\eta_\pi| < 1$) \cite{STAR_4pi}. 
The double-scattering contribution accounts only for $20\%$ of 
the cross section measured by the STAR Collaboration. 
The dash-dotted line represents a fit of the STAR Collaboration. 
Probably the production of the $\rho^0(1450)$ and  $\rho^0(1700)$ resonances
and their subsequent decay into the four-pion final state is 
the dominant effect for the limited STAR acceptance. 
Both, the production mechanism of
$\rho^0(1450)$ and $\rho^0(1700)$ and their decay into four charged
pions are not yet fully understood.  
A model for production of the resonances and their decay has 
to be work out in the future.
The right panel of Fig. \ref{Fig:dsig_dM4pi} shows four-pion invariant mass 
distribution for double-scattering mechanism for the limited range of 
pion pseudorapidity. The ALICE group collected
the data for four-charged-pion production with the limitation $|\eta_\pi|<1.2$.
We cannot compare this distribution with the ALICE data,
because those data points are not yet absolutely normalized.

\section{Conclusion}

We have studied two-$\rho^0$ as well as four-pion production 
in exclusive ultrarelativistic heavy ion UPC,
concentrating on the double-scattering mechanism of single-$\rho^0$
production.
The produced $\rho^0$ mesons decay, with almost $100\%$ probability,
into charged pions, giving large contribution to exclusive
production of the $\pi^+\pi^-\pi^+\pi^-$ final state.
We have compared contribution of four-pion production via
$\rho^0\rho^0$ production (double scattering and $\gamma\gamma$ fusion)
with experimental STAR Collaboration data.
The theoretical predictions have a similar shape 
as the distribution measured by the STAR Collaboration, but exhaust 
only about $20\%$ of the measured cross section. 
The missing contribution can come from decays of excited states 
of $\rho^0(770)$ into four charged pions.
In addition, predictions for the LHC have been shown. 


\end{document}